# *Ab Initio* Velocity-Field Curves in Monoclinic β-Ga₂O₃

Krishnendu Ghosh[†] and Uttam Singisetti[§]

Electrical Engineering Department, University at Buffalo, Buffalo, NY 14260, USA

[†]kghosh3@buffalo.edu, [§]uttamsin@buffalo.edu

**Abstract**: We investigate the high-field transport in monoclinic *β*-Ga₂O₃ using a combination of *ab initio* calculations and full band Monte Carlo (FBMC) simulation. Scattering rate calculation and the final state selection in the FBMC simulation use complete wave-vector (both electron and phonon) and crystal direction dependent electron phonon interaction (EPI) elements. We propose and implement a semi-coarse version of the Wannier-Fourier interpolation method [F. Giustino, M. L. Cohen, and S. G. Louie, Physical Review B, vol. 76, no. 16, 2007] for short-range non-polar optical phonon (EPI) elements in order to ease the computational requirement in FBMC simulation. During the interpolation of the EPI, the inverse Fourier sum over the real-space electronic grids is done on a coarse mesh while the unitary rotations are done on a fine mesh. This paper reports the high field transport in monoclinic β-Ga₂O₃ with deep insight on the contribution of electron-phonon interactions, and velocity-field characteristics for electric fields ranging up to 450 kV/cm in different crystal directions. A peak velocity of $2\times10^7$ cm/s is estimated at an electric field of 200 kV/cm.

## 1. INTRODUCTION

β-Ga$_2$O$_3$ is an emerging wide-bandgap semiconductor with excellent potential for power electronics applications. Experimentally demonstrated high breakdown voltage devices[1-5] makes it an attractive material for next generation power electronics in addition to deep UV optoelectronic applications[6-7]. However, the low symmetry monoclinic structure and the large primitive cell present immense challenges to accurate prediction of thermal, optical, and electrical transport properties. There have been a number of reports on thermal, dielectric, and optical properties of this material[8-15]. Recently, we reported low field electron mobility calculation in this material from first principles[16] that fairly agreed with experimental report[17]. However, there is no report on the high field transport properties in this material. High-field electron transport is important not just for electronics, but also for several branches of photonics. Electron phonon interactions (EPIs) play an important role in high-field transport. It is important to have a clear quantitative understanding of how EPIs affect high-field electron transport. Traditionally, they are modelled using phenomenological deformation potentials that could explain the experimental results. However, such approach ignores complex issues such as anisotropy and also requires experimental data to set the phenomenological constants. Revealing the right physics of EPI is crucial to engineer the electron and phonon dynamics. In recent years, there have been several studies on calculating phonon scattering rates in GaAs and Si[18-22] and low field mobilites[19,22] using EPI from *ab-initio* methods. However, including a full-band *ab-initio* based EPI in a Monte Carlo algorithm for high transport studies poses several computational challenges especially for materials with many phonon modes. Here, we systematically study the physics of EPI in β-Ga$_2$O$_3$ to reveal how they affect the high-field electron transport calculated using a full band Monte Carlo (FBMC) simulation and also address the computational challenges. Monoclinic β-Ga$_2$O$_3$ has a large primitive cell with low crystal symmetry which makes it an ideal benchmark system to study electron transport from first principles. The computational strategy is discussed first followed by the results obtained for β-Ga$_2$O$_3$.

## 2. THEORY AND METHODS

The computation begins with density functional theory (DFT) calculation on equilibrium lattice structure followed by the extraction of Kohn-Sham (KS) eigen values. Next, lattice response calculations are done based on the density functional perturbation theory (DFPT)[23]. This provides the dynamical phonon matrix and also the perturbation in the self-consistent potential for each perturbation. Diagonalizing the dynamical phonon matrix gives the phonon eigen values. The EPI elements are computed from the perturbation in the self-consistent potential[23]. The EPI elements and the phonon dynamical matrices are interpolated using a Wannier-Fourier interpolation scheme[24] to obtain fine sampling of the Brillouin zone. However, the long-range polar optical phonon (POP) EPI needs a separate treatment due to its aperiodic nature. Recent formulation[25] on long-range EPI interpolation scheme is used here.

### 2.1 Semi-coarse Electron-Phonon Interaction (EPI) interpolation

The fine resolution needed on both electron (**k**) and phonon (**q**) wavevector mesh requires large memory in computation which is exacerbated for materials with multiple phonon modes such as *β*-Ga$_2$O$_3$. Unlike scattering rate calculations, an 'on the fly' method[18,26] could not be used in the FBMC calculation since we need to store the EPI matrix elements for final state selection. An 'on the fly' calculation inside the final state selection program is inefficient since the summation over the real-space grid has to be performed for each electron, for each time interval, and for each scattering process making it extremely slow. Hence, it is necessary to store the fine grid matrix elements increasing the memory requirement. Here, we propose a semi-coarse Wannier-Fourier interpolation scheme to ease the computational requirement of storing the fine-grid matrix elements. In this method, the unitary rotations are on fine mesh but the inverse Fourier transform sum is on a coarse mesh. Storing the fine-grid electron-phonon interaction (EPI) elements require huge run-time memory. Using the definitions of electronic Wannier functions (WF)[24], we write the short range non-polar EPI matrix elements on fine **k** and **q** meshes as

$$g_{short}^v(\mathbf{k}, \mathbf{q}) = \frac{1}{N_e^2} U_{\mathbf{k}+\mathbf{q}} \left[ \sum_{\mathbf{R}_e, \mathbf{R}'_e} e^{i\mathbf{k}\cdot(\mathbf{R}_e - \mathbf{R}'_e)} g_{short}^v(\mathbf{R}_e - \mathbf{R}'_e, \mathbf{q}) \right] U_{\mathbf{k}}^\dagger \qquad (1)$$

Here, the unitary rotation matrices $U$ have a dimension of $N^w \times N^w$, where $N^w$ is the size of the Wannier subspace. $g_{short}^v(\mathbf{k}, \mathbf{q})$ is the short-range EPI on both fine meshes. $\mathbf{R}_e$ is the center of an unit cell in the supercell and $N_e$ is the total number of unit cells in the supercell. Eq.1 shows the inverse Fourier transform of the real-space (on electronic mesh) $g_{short}^v(\mathbf{R}_e - \mathbf{R}'_e, \mathbf{q})$ to an arbitrary **k**-point while the **q**-mesh interpolated information is already present on the right-hand side of the equation. The total memory requirement to store the short range EPI on both fine meshes would be governed by $N^k$ x $N^v$ x $N^q$ x $(N^w)^2$, which becomes large for the required $N^k$ and $N^q$ where $N^k$, $N^q$, and $N^v$ are the size of the fine *k*-mesh, *q*-mesh, and the number of phonon modes respectively. For this work the corresponding memory requirement becomes more than 500 terabytes. Hence, we propose a way to circumvent the problem. The primary contribution to $g_{short}^v(\mathbf{R}_e - \mathbf{R}'_e, \mathbf{q})$ comes when the two involved electronic WFs are on the same unit cell since the overlap of remote WFs are low (although not negligible and we discuss that later) in case of maximally localized WF (MLWF). Under such condition we can rewrite (after simple algebraic manipulation) Eq.1 as -

$$g_{short}^{v,0}(\mathbf{k}, \mathbf{q}) = \frac{1}{N_e} U_{\mathbf{k}+\mathbf{q}} g_{short}^v(\mathbf{0}_e, \mathbf{q}) U_{\mathbf{k}}^\dagger = U_{\mathbf{k}+\mathbf{q}} U_{\mathbf{k}_c+\mathbf{q}}^\dagger g_{short}^{v,0}(\mathbf{k} = \mathbf{k}_c, \mathbf{q}) U_{\mathbf{k}_c} U_{\mathbf{k}}^\dagger \qquad (2)$$

Hence the same unit cell contribution $g_{short}^{v,0}(\mathbf{k}, \mathbf{q})$ can be accessed just by storing $g_{short}^{v,0}(\mathbf{k} = \mathbf{k}_c, \mathbf{q})$ on a coarse **k**-mesh ($\mathbf{k}_c$) and the rotation matrices. This reduces the memory requirement by several orders. Next, we discuss the contribution from remote WFs. Now we consider Eq.1 with $\mathbf{R}_e \neq \mathbf{R}'_e$. Under such conditions, a reduction similar to Eq.2 would have been perfectly possible if the phase factor $e^{i\mathbf{k}\cdot(\mathbf{R}_e - \mathbf{R}'_e)}$ was absent. If the unit cells are far enough from each other (*e.g.* $\mathbf{R}_e - \mathbf{R}'_e > 5\times$ lattice spacing), $g_{short}^{v}(\mathbf{R}_e - \mathbf{R}'_e, \mathbf{q})$ becomes negligible. However, if the unit cells are close to each other, because of the **k**-space smoothness of the matrix elements in Wannier gauge the contribution from the remote unit cells are smaller compared to the same unit cell. Under such a gauge-choice, treating the remote WF contribution similar to Eq.2 introduces small inaccuracy in calculation but eases the computational requirement to a great extent. So the overall $g_{short}^{v}(\mathbf{k}, \mathbf{q})$ can be obtained from $g_{short}^{v}(\mathbf{k} = \mathbf{k}_c, \mathbf{q})$ and the rotation matrices on the fine mesh. The essential idea is to rotate the coarse-mesh elements to an optimal Bloch-space that produces the MLWFs, perform Fourier transform to real space followed by inverse Fourier transform on another coarse **k**-mesh. The fine **k**-mesh elements could be obtained 'on the fly' just by un-rotating the elements back to the normal Bloch-space using the fine **k**-mesh rotation matrices. This reduces the 'on the fly' time consumed in converting the elements back and forth between reciprocal and real spaces while at the same time mitigating the huge memory requirement for storing the elements on both fine **k** and **q** meshes. It is important to mention here that this sacrifice of accuracy does not lead to any sacrifice of the crystal symmetry mediated selection rules of non-polar scattering since that is taken care of inside short range EPI matrix $g_{short}^{v}(\mathbf{R}_e - \mathbf{R}'_e, \mathbf{q})$.

## 2.2 Full Band Monte Carlo (FBMC) simulation

High-field transport requires Monte Carlo (MC) simulations. The method is widely used for velocity field curves, we briefly describe the method[27,28,30,31] here. We solve the Boltzmann transport equation (BTE) with FBMC technique starting from *ab initio* KS eigen values on a fine reciprocal **k**-mesh, phonon eigen values, both long-range (POP) and short range (non-polar) EPIs, and the scattering rates on the fine **k**-mesh with contributions from each phonon modes treated separately. For the phonon eigen calculations we used two different types of **q**-meshes one for short-range elements that span the entire Brillouin zone (BZ), while the other for the long range elements which exist only in the vicinity of the zone center. Our MC scheme is spatially homogenous and hence the spatial gradient term in the BTE disappears. The ensemble of electrons is first initialized in a Maxwellian thermal distribution before the electric field is turned on. Electrons are allowed to drift under the influence of the electric field and to get scattered randomly. The free drifting of the electron changes the crystal momentum according to (in atomic units) $\dot{\boldsymbol{k}} = \boldsymbol{F}$ where $\boldsymbol{F}$ is the applied electric field. The free drifting time ($t_0$) is estimated[28] based on a random number $r$ such that, $t_0 = -\frac{1}{\Lambda}\ln r$ where $\Lambda$ is the maximum possible scattering rate from a given band at any $\boldsymbol{k}$ point. The final state selection after a scattering mechanism is described next.

The first step of the final state selection process is to identify the mechanism that caused the scattering. A 'mechanism' includes the details of phonon mode index, final electronic band index, and the nature (polar/non polar, absorption/emission) of the scattering. So, for 30 phonon modes and 2 electronic bands, the allowed number of mechanisms would be 180. It is noted that polar mechanisms do not contribute to interband scatterings. Finding the scattering mechanism follows from a normalized scattering rate (*NSR*) table that is formed at each **k** point and each band from the previously computed *ab initio* scattering rates. The formation of the NSR table is discussed here briefly. If $S_m^v(\mathbf{k})$ denotes the scattering rate of an electron at a wave-vector **k** and an initial band *m* mediated by a mechanism $v$. The elements of the NSR table at band *m* and wave-vector **k** has a form –

$$NSR(v) = \frac{\sum_{i=1}^{v} S_m^i(\mathbf{k})}{max_k \sum_{i=1}^{N_m} S_m^i(\mathbf{k})} \qquad (3)$$

So the NSR elements always lies in (0, 1) and NSR is non decreasing with increasing $v$. The NSR elements are stored in a look-up table whose memory requirement scales as $N^k \times N^b \times N^m$, where $N^b$ is the number of bands taken in the transport calculation and $N^m$ is the total number of scattering mechanisms. The large memory requirement of the NSR in a parallel computing environment is met by storing it on shared memory windows[29] available to all processes running under the same computing node. While selecting the mechanism a random number $r'$ [in (0, 1)] is generated and the mechanism $v$ is selected if the relation $NSR(v) > r > NSR(v-1)$ holds true. If $NSR(N^m) < r$, the electron is 'self-scattered'[27,28,30,31] which means the state of the electron remains unchanged.

Given an 'actual' scattering event occurs, once the phonon mode, final electronic band, and the nature of the scattering are extracted from the *NSR* table, the next task is to find the final electronic wave-vector. We use a method tactically similar to the one reported for covalent semiconductors[29] and we treat the polar phonon mode scatterings separately. Here a brief details of the algorithm is given. Let us consider the initial electron band index is *m* and wave-vector is $k_i$. First, all the **k**-points $k_j$ in the final band *n* that satisfy the energy conservation are shortlisted. Next, a further shortlisting is done based on the strict energy and momentum conservation with implementation of phonon dispersion. So this gives a list of final $k_j$ points that satisfy both the conservations. Now we exploit the full-band EPI elements and this is where both **k** and **q** dependence of the EPI elements are required. Another normalized table, (similar to Eq. 3) is formed based on the product of the local density of states (LDOS) at the potential $k_j$s and EPI strength at the corresponding momentum conserving phonon wave-vector, $q_j$. This is where the polar scattering needs to be taken care of separately. The small-angle preference of the polar mechanisms needs to be properly captured in order to reflect the lower momentum relaxation time compared to corresponding energy relaxation time. This is done with the aid of a separate fine **q**-mesh that exists only near the zone center. Using a random number the normalized table is scanned through like it was done for NSR and a final $k_j$ is selected. However, this only gives the grid point in the reciprocal space. The final electron wave-vector could belong to anywhere within the small cube that is represented by this grid point. The scheme to select the final wave-vector within the small cube is exactly same as the one described in steps (e-g) of Fig. 7 in Ref. 29. Next we discuss the calculation and results for β-Ga$_2$O$_3$.

3.   **RESULTS AND DISCUSSION**

We carry out DFT calculations on *β*-Ga$_2$O$_3$ as described in our previous work[16] with similar pseudo-potentials[35] and zone sampling[36] using Quantum ESPRESSO[37]. Fig 1(a) shows the crystal structure of *β*-Ga$_2$O$_3$ along with the Cartesian direction convention that is followed throughout this work. In Fig. 1(b) we show the resulting KS eigen values along two reciprocal vectors. Four conduction bands are shown out of which first two are used in subsequent transport calculations. Fig. 1(c) shows the equi-energy surfaces for two different energies. It can be seen that at a lower energy the surface is spherically symmetric, while at higher energies the surface becomes anisotropic with contributions from higher energy bands. We interpolated the KS Hamiltonian through MLWFs by using the wannier90 code[38]. Our optimal Wannier subspace (S) consists of two low lying conduction bands within a frozen energy window. The phonon calculation is done under DFPT using QE. The phonon eigen values and the short-range (non-polar) EPI matrix elements are interpolated on a fine **q**-mesh using a modified version of the EPW code[26].

### 3.1 EPI in $\beta$-Ga$_2$O$_3$

The short-range non-polar EPI elements are primarily responsible for higher momentum and energy relaxation rates at high-fields that lead to saturation in velocity. The Wannier-Fourier interpolated short-range matrix elements are shown in Fig.2 (a) with respect to **q**. We show the sum of the squared magnitude of the coupling over all the vibrational modes. The long-range POP coupling is also shown on the same plot for comparison. The **q** dependence of EPI matrix elements ($\sum_v |g_{short}^v|^2$) is to be noted along with its crystal direction dependence ($\Gamma$-Z and $\Gamma$-N), the strength is minimized at the $\Gamma$ point. The inset shows a contour plot on the 2D plane formed by the same two directions. The symmetry of the short-range element observed on the contour plot is a signature of the inversion symmetry associated with the C$_{2m}$ space group[39]. The small discontinuity on the short-range EPI marked by the arrow is a result of the gauge ambiguity arising from degenerate final electronic states (see Fig 1(b)). Such discontinuity does not affect the total scattering rate since the latter is gauge-invariant. Fig. 2(b) shows the short-range EPI for each phonon mode for two different **q** vectors. The red bars show $|g_{short}^v|^2$ for initial electron wave-vector **k** at 0.25×$\Gamma$Z away from $\Gamma$ while the phonon wave-vector (**q**) is 0.5×$\Gamma$Z away from $\Gamma$. The corresponding two points from the green bars are 0.125×$\Gamma$N (**k**) away from $\Gamma$ and 0.25×$\Gamma$N (**q**) away from $\Gamma$. The inset of Fig. 2(b) shows the scattering schematic for high energy electrons. The long arrows correspond to a direction reversal of an electron which provides a high momentum relaxation. For high-field transport the available phase space allows phonon wave-vectors from all directions. But the higher momentum relaxing phonon wave-vectors need to be inclined as much opposite as possible to the propagating electron wave-vector. This explains the choice of electron and phonon wave-vectors to show the individual mode contributions in Fig.2 (b). It could be seen that for an electron with initial wave-vector at the Γ point the dominating non-polar phonon modes are at 83 meV and 96 meV along Γ-Z, while along Γ-N they are 22 meV and 90 meV. Fig. 2(c) shows the short range EPI element ($g^v_{short}(\mathbf{k})$) along the Γ-Z-Γ direction calculated using coarse mesh, fine mesh and the semi-coarse sampling proposed in this work. The **q** vector for this calculations is kept fixed at halfway along the ΓZ. The DFPT calculated coarse-mesh element is shown with red dots while a regular fine mesh interpolated result is shown with red solid curve. The kinks on the semi-coarse curves occur when the representative coarse **k**-mesh point corresponding to a fine-**k** mesh point changes. The small errors incurred due to this strategy is acceptable given the resulting ease in computational memory requirement. The key aspect is that the coarse mesh used in the Fourier transform does not have to be as coarse as the original DFT mesh. It should be made as fine as possible within the available computational resources. The maximum observed deviation between the 8×8×4 semi-coarse mesh and the fine mesh elements is ~ 12% along the Γ-Z-Γ direction for this given **q** vector.

**3.2 Scattering rates and high-field transport**

Next we discuss the Fermi-Golden rule computed non-polar phonon scattering rates. The energy conserving delta function is implemented using a Lorentzian smearing. We carried out an analytical fitting of the computed scattering rates using a deformation potential (DP) approximation[31] to guide future analytical transport calculations. However, it is important to note that resulting fitting parameters do not provide any actual physical picture of EPI or existence of phonon modes. The density of states used in the scattering rate fitting procedure is computed from the *ab initio* band-structure. For non-polar optical phonon scattering (Fig. 3(a)) we fitted the computed scattering rates with $D_0^2 / \omega_0 = 7.2 \times 10^{18}$ eV/cm$^2$, where, $D_0$ is the optical deformation potential and $\omega_0$ is the optical phonon energy. For a suitable fitting, these two parameters are inter-adjustable and hence are not unique. Fig 3 (a) shows the fitted curves with dashed curves. Fig. 3(b) shows the calculated acoustic phonon scattering rate with energy. For acoustic phonons the fitting procedure is two-fold as seen in Fig. 3(b). Scattering due to the zone-center (ZC) phonons are fitted with $D_A^2/v_s^2 = 5\times10^{-11}$ eV$^2$s$^2$/cm$^2$, where $D_A$ is the acoustic deformation potential and $v_s$ is velocity of sound. On the other hand, the zone-edge (ZE) acoustic phonons have a dispersion similar to optical phonons, hence we use an optical phonon like fitting with $D_{AZE} = 5\times10^7$ eV/cm and $\omega_{AZE} = 0.01$ eV (dashed green lines in Fig. 3 (b)). A single fitting of the acoustic scattering rate is not possible because of the approximations[31] used to derive the analytical form of the scattering rate namely, the negligible phonon energy and the linear dependence of the coupling with phonon wave-vector.

Now, we discuss the results of the Monte Carlo simulation. Fig. 4 (a) shows the initialized Maxwell electron distribution in energy space at room temperature at the start of the FMBC simulation, while Fig. 4(b) shows the initial electron distribution along $k_z$ in the direction (z-direction) of the applied electric field. The respective bottom panels (Fig. 4 (c) and Fig. 4 (d)) show similar plots for an applied electric field of 300 kV/cm along the *z* direction. It is noted that the non-equilibrium distribution in energy space die off well beyond 2 eV of energy. The first satellite valley occurs at ~ 2.5 eV, hence very few electrons are transferred to satellite valleys at the field strength. That is further supported by the electron distribution as a function of $k_z$ seen In Fig. 4(d), which shows very low population near the zone edge where the valley occurs. Next, we discuss the transient response of the drift velocity as shown in Fig. 5. At low electric fields the transport is dominated by POP scattering and the momentum relaxation rate is not increasing with electron energy implying the drift velocity monotonically arrives to a steady state. However beyond 150 kV/cm velocity overshoot starts to appear. This is attributed to the fact that with increasing electron energy the intra-band non-polar scattering becomes significantly high allowing short-range transitions that boost up momentum relaxation drastically while the energy relaxation is still limited by the energy of the phonon modes. The imbalance of the two rates results[31] in a slow rise of the electron energy than the same in the drift velocity. However, the momentum relaxation rate is also dependent upon the electron energy and hence it gets adjusted in a larger time-scale than that of the drift velocity and hence we see the overshoot.

Fig. 6 shows the velocity-field curves in three different Cartesian directions calculated from FBMC using the scattering strengths discussed in the previous sections. We see that the velocity increases in all three directions up to 200 kV/cm followed by a negative differential conductivity (NDC). The calculated NDC is less than that observed in GaAs since the satellite valleys occur at a much higher energy (the lowest one being at ≈2.4 eV) as compared GaAs (≈0.3eV). Unlike GaAs and GaN where the NDC results from inter-valley scattering, the NDC and velocity saturation in $\beta$-$Ga_2O_3$ results from intra-valley short-range scattering. In $\beta$-$Ga_2O_3$ the non-parabolic nature of the $\Gamma$ valley at higher electronic energies reduce the average electronic group velocities resulting in the NDC and the short range intra-valley EPI results to the saturation of velocity. For the range of electric fields considered here, electrons barely reach the satellite valleys (See Fig. 4(c)). So the fact that the velocity starts rolling off from its peak value beyond 200 kV/cm is to be attributed to the non-parabolicity of the conduction band rather than intervalley scatterings. The average peak velocity at an electric field of 200 kV/cm is ≈2 × $10^7$ cm/s which is slightly lower than wurzite GaN[40]. Next, we discuss the observed anisotropy in the velocity-field curves. Prior to the peak velocity point, around 100 kV/cm the *z* direction velocity is lower than that in *x* and *y* directions. This is attributed to the large polar optical phonon scattering by the $B_u^1$ mode which is polarized in the *z* direction as was observed in our previous work[16]. More details on this anisotropy could be found in a future work[41]. However, beyond the peak velocity the drift velocity is relatively higher in the *y* direction. We attribute this to the fact that the slope of the energy band drops at a higher **k** in the *y* direction compared to that in the other two directions. Hence the ensemble average of the drift velocity is higher in the *y* direction. The velocity-field curves were fitted using Barnes model[42] that formulates NDC with, $\mu_n(F) = \frac{\mu_0 + \frac{v_{sat}}{F}\left(\frac{F}{F_c}\right)^\gamma}{1+\left(\frac{F}{F_c}\right)^\gamma}$, where $\mu_0$ is the low-field mobility and $v_{sat}$, $F_c$, and $\gamma$ are adjustable fitting parameters. This model is used in commercial device simulators[43] to simulate device operation. The fitting parameters in three different directions are given in Table-I. It is noted that, these parameters are good only upto an electric field of about 500 kV/cm. Beyond that other physical phenomena such as interband transitions and impact ionization are likely to kick in which will modify the velocity-field curves.

## 4. SUMMARY

In summary, we have highlighted the role of full-band EPI in controlling the electron transport in β-Ga$_2$O$_3$. A semi-coarse **k**-mesh version of the Wannier-Fourier interpolation[24] strategy is proposed for the short-range EPI in order to make it computationally tractable in a Monte Carlo algorithm. The applicability of this strategy is valid within the current scope of this work, however, for validating this method for arbitrary material systems, further study is required which itself could be a topic for future work. The calculated EPIs are used in an FBMC algorithm to analyze high-field transport. Scattering rate fitting parameters are calculated to guide analytical calculations. Monoclinic β-Ga$_2$O$_3$ is studied under this theoretical manifold followed by the prediction of the velocity-field characteristics. Velocity-field curves are fitted in different directions using compact NDC models to guide device design.

The authors acknowledge the support from the National Science Foundation (NSF) grant (ECCS 1607833). The authors also acknowledge the excellent high performance computing cluster provided by the Center for Computational Research (CCR) at the University at Buffalo.

**FIGURE CAPTIONS**

**Fig. 1**: (a) Crystal structure of $\beta$-Ga$_2$O$_3$. $a$, $b$, and $c$ represent the conventional lattice directions for monoclinic crystals. The Cartesian $x$, $y$ and $z$ direction used for the calculations is also shown. (b) The DFT calculated conduction band energies are shown for $\beta$-Ga$_2$O$_3$. The first two bands (shown in red and blue) are used in the transport calculation. (c) The equi-energy surfaces are shown for two different energies. It is evident that at lower energy the surface is spherically symmetric while higher energy introduces band anisotropy and also multiple bands. (a) is visualized by Vesta[32, 33] and (c) is visualized by XCrySDen[34].

**Fig. 2**: (a) The dependence of the EPI elements with the phonon wave-vector (**q**). The inset shows a contour plot of the short-range EPI on the $\Gamma$Z-$\Gamma$N plane. The initial electronic state is taken at the $\Gamma$ and hence the inversion symmetry of the crystal is reflected in the contour plot. (b) The mode-wise splitting of the short-range EPI for two **q** points along two different directions. The phonon energies (meV) for the dominating modes are shown with corresponding colors. (Inset) At higher electron energies a high momentum relaxation requires phonon wave-vectors that are inclined towards opposite direction of the electronic wave-vector. (c) The interpolated elements in **k**-space using the semi-coarse **k**-mesh strategy. The **q** point is kept fixed at a point halfway along $\Gamma$Z. A pure fine-mesh interpolation is also shown for comparison. In (a-c) the initial and final electronic band indices are taken to be the first band.

**Fig.3**: (a) Comparison of computed non-polar optical phonon (NOP) scattering rate and fitted NOP scattering rate. (b) Comparison of computed acoustic phonon scattering rate with the corresponding fittings. The fitting splits the contribution from zone-center and zone edge acoustic modes. See text for fitting parameters.

**Fig. 4:** Electron population in energy space and $k_z$ space. (a-b) show the histogram plots for intial conditions used in the FBMC simulation and (c-d) show the corresponding plots for steady state under an external electric field of 300 kV/cm applied along the *z* direction.

**Fig. 5:** Transient response of drift velocity for different electric fields.

**Fig. 6:** Velocity-field characteristic of β-Ga$_2$O$_3$ at room temperature in three different directions.

**Table –I:** Analytical fitting parameters for the computed velocity-field curves

**Fig. 1**:

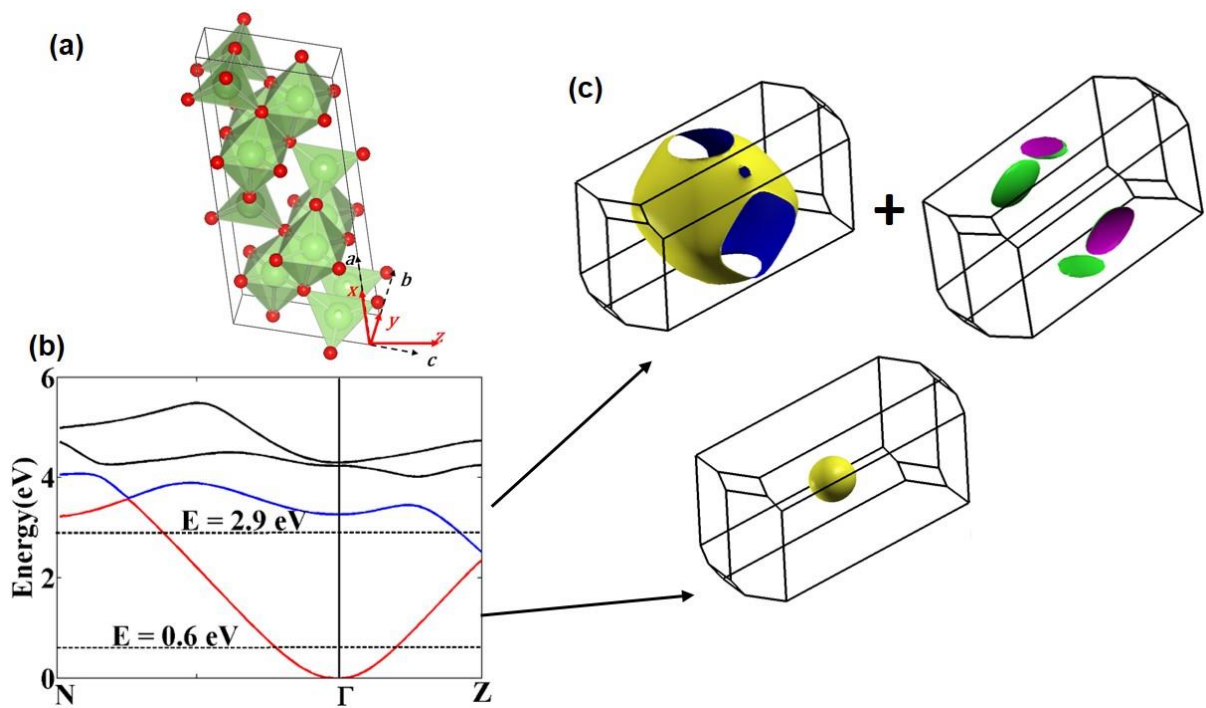

**Fig. 2**

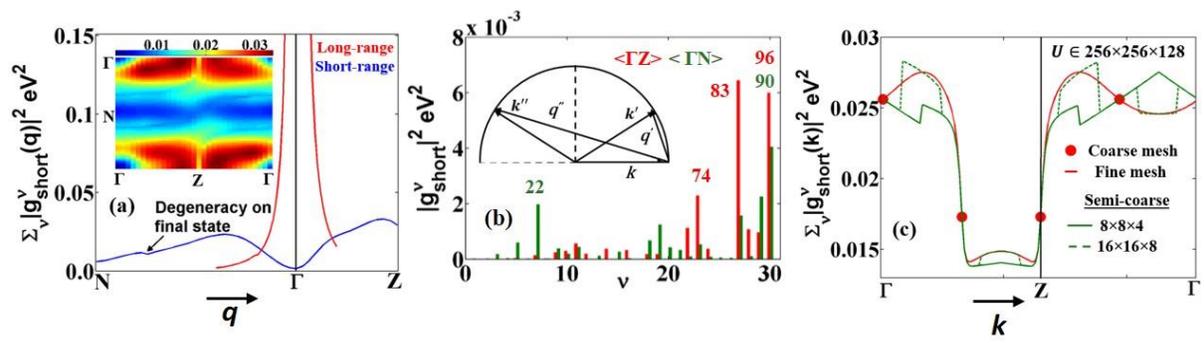

Fig. 3

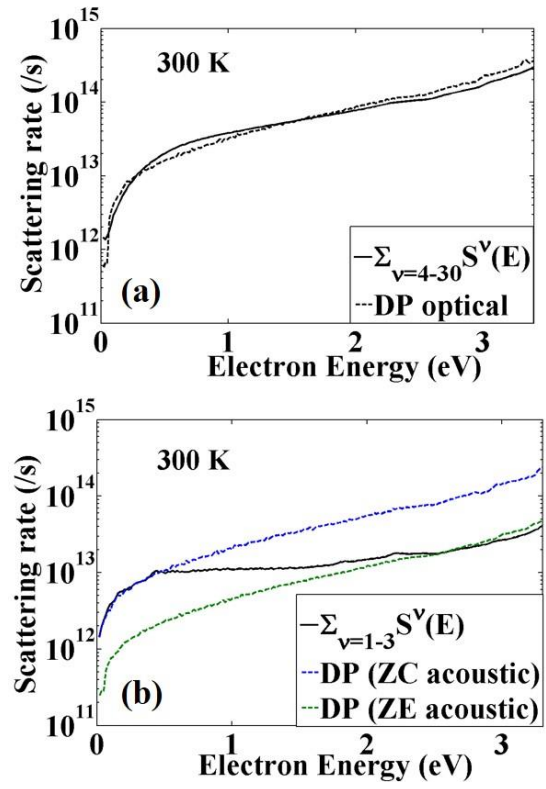

**Fig. 4**

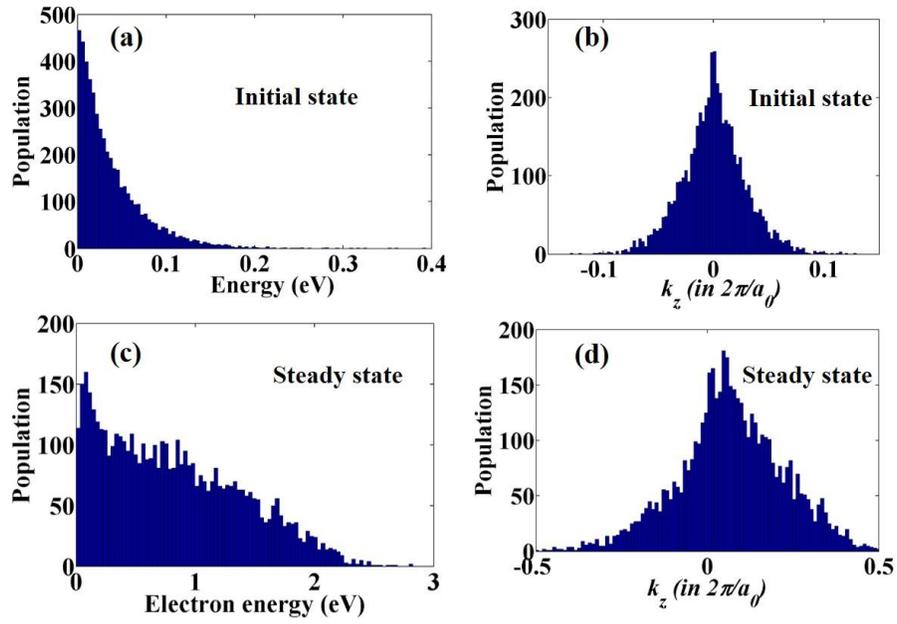

**Fig. 5**

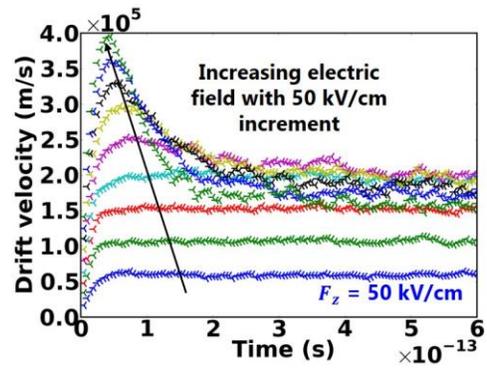

**Fig. 6**

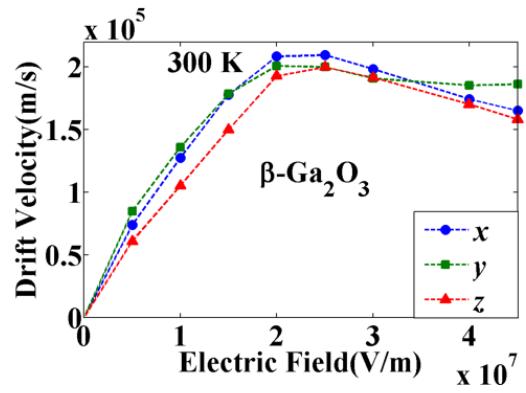

**Table –I**

|  | $x$ | $y$ | $z$ |
| --- | --- | --- | --- |
| $\mu_0$ (cm$^2$/Vs) | 140 | 140 | 112 |
| $v_{sat}$ (cm/s) | $10^7$ | $1.5 \times 10^7$ | $10^7$ |
| $F_c$ (V/cm) | $2.25 \times 10^5$ | $1.54 \times 10^5$ | $2.63 \times 10^5$ |
| $\gamma$ | 2.84 | 2.47 | 3.35 |